\documentclass[prl,twocolumn,amsmath,amssymb,superscriptaddress,showpacs]{revtex4-1}

\usepackage{amsmath,amssymb,graphicx}
\input epsf
\usepackage{amsbsy}
\usepackage{latexsym}
\usepackage{color}
\usepackage{graphicx}
\usepackage{psfrag}
\usepackage[normalem]{ulem}
\usepackage{bm}

\newcommand{\be}{\begin{equation}}
\newcommand{\ee}{\end{equation}}
\newcommand{\bea}{\begin{eqnarray}}
\newcommand{\eea}{\end{eqnarray}}
\newcommand{\Qpca}{Q_{\mbox{\tiny PCA}}}
\newcommand{\comment}[1]{}

\begin{document}

\title{Dynamical freezing of relaxation to equilibrium}

\author{Stefano Iubini}
\email{stefano.iubini@unipd.it}
\affiliation{Department of Physics and Astronomy, University of Padova, Via Marzolo 8, I-35131 Padova, Italy}
\affiliation{Istituto dei Sistemi Complessi, Consiglio Nazionale delle Ricerche, Via Madonna del Piano 10, 50019 Sesto Fiorentino, Italy}

\author{Liviu Chirondojan}
\affiliation{SUPA and Department of Physics, University of Strathclyde, Glasgow G4 0NG, Scotland, UK}

\author{Gian-Luca Oppo}
\affiliation{SUPA and Department of Physics, University of Strathclyde, Glasgow G4 0NG, Scotland, UK}

\author{Antonio Politi}
\affiliation{Institute for Complex Systems and Mathematical Biology \& SUPA
           University of Aberdeen, Aberdeen AB24 3UE, Scotland, UK}

\author{Paolo Politi}
\affiliation{Istituto dei Sistemi Complessi, Consiglio Nazionale delle Ricerche, Via Madonna del Piano 10, 50019 Sesto Fiorentino, Italy}
\affiliation{INFN Sezione di Firenze, via G. Sansone 1, 50019 Sesto Fiorentino, Italy}

\begin{abstract}
We provide evidence of an extremely slow thermalization occurring in the Discrete NonLinear Schr\"odinger (DNLS)
model. At variance with many similar processes encountered in statistical mechanics - typically ascribed 
to the presence of (free) energy barriers - here the slowness has a purely dynamical origin: 
it is due to the presence of an adiabatic invariant, which freezes the dynamics of a tall breather.
Consequently, relaxation proceeds via rare events, where energy is suddenly released towards the background.
We conjecture that this exponentially slow relaxation is a key ingredient contributing to the non-ergodic behavior recently observed in
the negative temperature region of the DNLS equation.
\end{abstract}

\maketitle

Statistical physics offers several examples of slow processes. In many cases the existence of long time scales can be traced back to the presence of (free) energy barriers, which require the emergence of strong fluctuations for them to be overcome. Structural and spin glasses, as well as colloids are strongly affected  by this mechanism,~\cite{berthier2011theoretical,Cipelletti2005} where frustration and disorder can either give rise to aging phenomena~\cite{Cugliandolo2002} and jamming~\cite{trappe2001jamming} or to ergodicity breaking~\cite{palmer1982broken}, if they become  insurmountable in the thermodynamic limit.
Free-energy barriers can also be dynamically induced by kinetic constraints~\cite{sollich1999,garrahan2002} in pure, unfrustrated systems.
Slow phenomena can however emerge in the absence of free-energy barriers, if the onset of equipartition is slowed down due
to  phase-space regions characterized by a nearly integrable dynamics~\cite{fermi1955,berman2005fermi,gallavotti2007fermi}.
\comment{In this manuscript we consider another different example of extremly slow evolution whose origin is dynamical rather than energetical.}

\comment{
Frozen dynamics also appears in driven systems, e.g. in active matter~\cite{Reichhardt2011} 
if the motion appears to be confined
to a specific region, and in quantum systems~\cite{balz2017typical,lan2018quantum}. 
Another set-up which  has recently attracted the interest of scientists for an unusual transport regime 
is a system of rotors, 
where the conductivity is exponentially small upon increasing the temperature~\cite{pino16}.  
This phenomenon has been shown to be related to the emergence of many-body 
quantum localization and ergodicity breaking. 
Moreover, rigorous results
obtained for small rotor chains in contact with external reservoirs prove that relaxation 
to nonequilibrium stationary states involve stretched-exponential 
rates~\cite{Cuneo_3rotators2015,cuneo2016_4r,cuneo2017relaxation}.
}

In the Discrete NonLinear Schr\"odinger (DNLS) equation, the subject of this Letter,
slow phenomena may emerge as a result of intrinsic localized fast rotations usually referred to as {\it discrete breathers}~\cite{chris2003discrete,ng2009,Franzosi2011,iubini2013,eckmann2017}. 
This equation, which models propagation in nonlinear discrete media with 
negligible dissipation~\cite{Holstein,Eilbeck1985,Kevrekidis}, is used to study many physical applications: 
trapped ultra-cold gases~\cite{trombettoni2001,livi2006,hennig2010transfer}, magnetic systems~\cite{borlenghi14,borlenghi15} and arrays of optical wave-guides~\cite{jensen1982,christodoulides1988}.

Its fame is also due to the so-called negative-temperature region~\cite{Rasmussen2000,iubini2013,levy2018equilibrium,cherny2018non}, where equipartition is violated due to the spontaneous emergence of breathers out of a noisy background.
Statistical-mechanical arguments~\cite{Rumpf2004,rumpf2007,rumpf2008,rumpf2009} show that the density of breathers should progressively decrease until a final state is reached where a single breather collects the excess energy from the background.
Such relaxation process, induced by purely entropic forces, 
 has been understood to be  a condensation phenomenon~\cite{szavits2014,Evans_Majumdar_condensation2014,barre18}
due to the existence of two conserved quantities, the mass and the energy.
However, the simplest condensation models~\cite{iubini2013,IPP14,iubini17} yield
 to a power-law coarsening of breathers, while 
microcanonical simulations of DNLS dynamics give evidence of a quasi-stationary regime where the number of breathers fluctuates around a well-defined average value, implying that ergodicity is broken~\cite{iubini2013,mithun18}. 
\comment{Furthermore, nonequilibrium grandcanonical simulations suggest the occurrence of 
negative-temperature regimes where high breathers do not even arise~\cite{wall17}.
}

The goal of this Letter is to clarify the nature of the slow processes observed in the DNLS equation.
First, we show that slow breather dynamics appears in the positive temperature regime too if we prepare the system with a tall breather sitting on a noisy background and let it relax. Second, we give evidence that relaxation is slow because an adiabatic invariant blocks diffusion, thereby leading to the effective ergodicity breaking discussed in Ref.~\cite{mithun18}.

The DNLS equation has the form
\begin{equation}
i \dot {z}_n = -2 |z_n|^2z_n - z_{n+1}-z_{n-1} \; ,
\label{eq:dnls}
\end{equation}
where $z_n$ are complex variables, $n=-N_0,\dots,N_0$ is the index of the lattice site and 
open boundary conditions are assumed.  The model has two exactly conserved quantities, namely the total energy
\begin{equation}
 H= \sum_{n=-N_0}^{N_0} \left( |z_n|^4+z_n^*z_{n+1}+z_nz_{n+1}^* \right) ,
 \label {Hz}
 \end{equation}
and the total mass, $A=\sum_n |z_n|^2$, related to the invariance under time translation ($t\to t+\bar t$) and phase rotation ($z_n \to z_n e^{i\bar\phi}$), respectively. 
If $h=H/N$ and $a=A/N$ (with $N=2N_0+1$) are the density of energy and mass, respectively,
the curve $h=2a^2$ defines the equilibrium states at infinite
temperature $T=(\partial s / \partial h|_a)^{-1}$,
$s$ being the entropy density~\cite{Rasmussen2000,iubini2013}.

In Eq.~(\ref{eq:dnls}) breathers naturally appear 
in the $T<0$ region (defined by $h>2a^2$~\cite{Rasmussen2000}): 
their dynamics looks essentially frozen~\cite{iubini2013}. 
Conversely, for $T>0$ (i.e. $a^2-2a<h<2a^2$~\cite{Rasmussen2000}), 
breathers are entropically disadvantaged and must decay~\cite{Rasmussen2000,Rumpf2004,rumpf2007,rumpf2008,rumpf2009}.
In this Letter we probe the positive-temperature frozen dynamics by studying the relaxation of a single breather initially set at $n=0$, 
with $T$ and the chemical potential $\mu$ imposed by external reservoirs acting on both chain ends~\cite{sup_mat02}.

Breather stability has been already studied in the literature, but exclusively in the presence of
a weakly fluctuating (small-amplitude) background~\cite{johansson2004}.
Here, we consider a generic-amplitude background, which cannot be treated perturbatively.

\begin{figure}
\begin{center}
\includegraphics[width=0.45\textwidth,clip ]{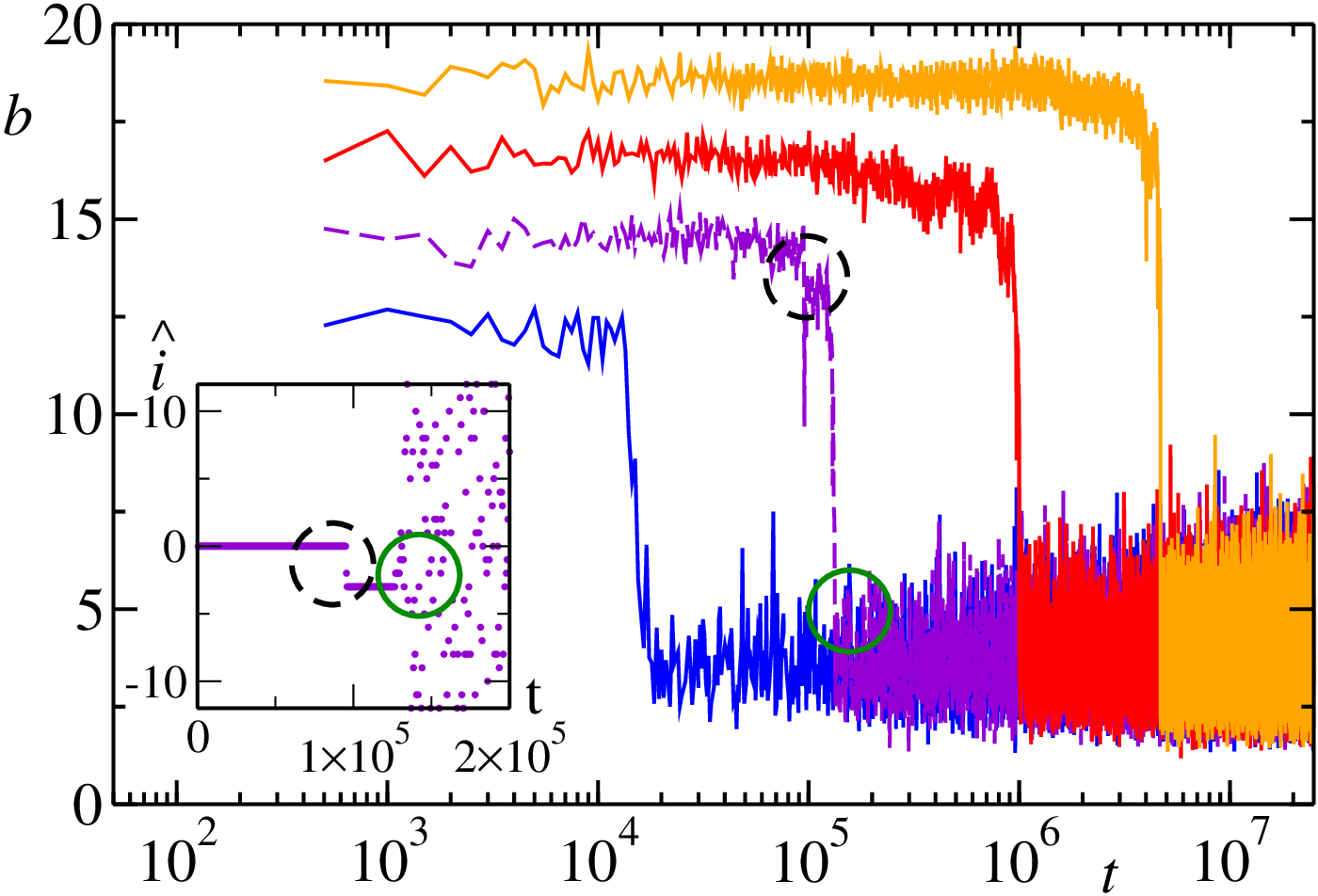}
\end{center}
\caption{Relaxation of the breather mass $b$ on a positive temperature background for increasing initial breather heights in a 
DNLS chain with $N=31$ in contact with two thermal baths at $T=10, \mu=-6.4$. 
The inset shows the breather position $\hat{i}(t)$ for the violet dashed line. 
Dashed black and full green circles identify respectively a jump and the final destruction of the breather, i.e. the onset of equipartition.
}
\label{fig_brt_relax}
\end{figure}

Exemplary traces of the evolution of the mass $b(t) = |z_0^2(t)|$ for a background temperature $T=10$ and different 
$b(0)$ values are plotted in Fig.~\ref{fig_brt_relax}. There, we notice a dramatic increase of the lifetime with
the initial mass (the horizontal scale is logarithmic).

\begin{figure}
\begin{center}
\includegraphics[width=0.45\textwidth,clip ]{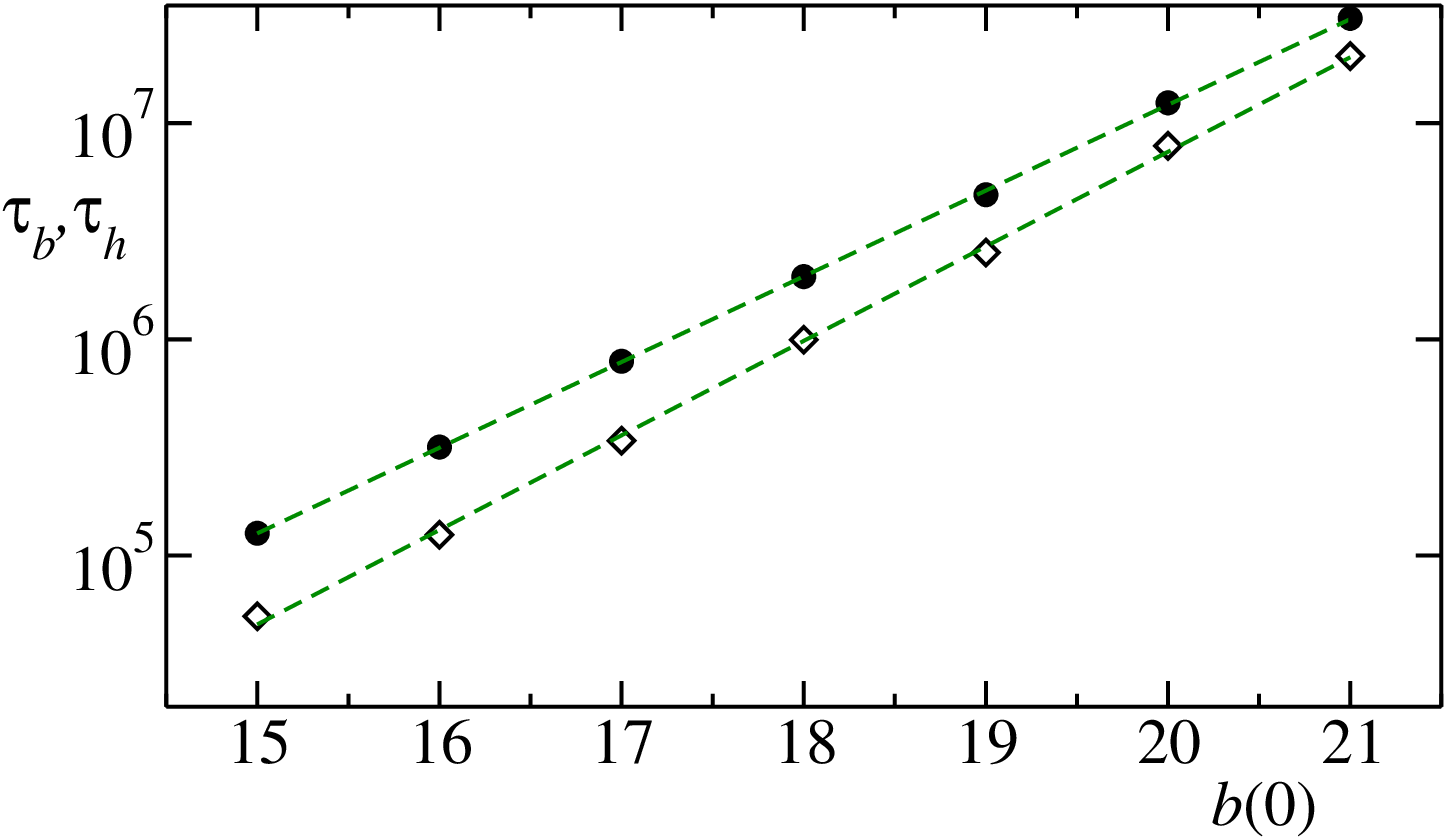}
\end{center}
\caption{Breather lifetime $\tau_b$ (black dots) and first hopping time $\tau_h$ 
(empty diamonds) for a chain with $N=31$, $T=10$ and chemical
potential $\mu=-6.4$ versus the initial breather mass $b(0)$. Relaxation times are
computed by the first passage time, averaging over 100 realizations.
Symbol sizes are of the order the standard deviations of lifetimes.
}
\label{fig_lifetime}
\end{figure}

The dependence of the average lifetime $\tau_b$ on the initial mass is analyzed in a more quantitative way
in Fig.~\ref{fig_lifetime}~\cite{sup_mat02}. In each single realization, $\tau_b$ is determined as the
shortest time such that $b(t) \le \theta$, where $\theta$ is a suitable threshold. As we are 
interested in tall breathers ($b(0)\gg 1$), and given that $b(t)$ is characterized by a fast final drop
(see Fig.~\ref{fig_brt_relax}), the choice of $\theta$ is not a critical issue~\cite{sup_mat02}.
We find that $\tau_b$ increases as $\tau_b \approx \mathrm{e}^{\alpha b(0)}$, with an exponent 
$\alpha = 0.91 \pm 0.01$ (see black dots in Fig.~\ref{fig_lifetime}).

In the following, we discuss the origin of this scaling behavior, starting from the empirical observation
(see Fig.~\ref{fig_brt_relax}) that the mass evolution is characterized by seemingly stationary ``laminar" 
periods~\footnote{The increasing size of the fluctuations after breather death 
is a deceptive effect due to the horizontal logarithmic scale, 
which compresses an increasing number of points in the same interval at longer times.}
accompanied by a few localized episodes, where the breather amplitude decreases abruptly 
(though, in some cases, upward jumps are observed as well).

\begin{figure}[ht]
\begin{center}
\includegraphics[width=0.45\textwidth,clip ]{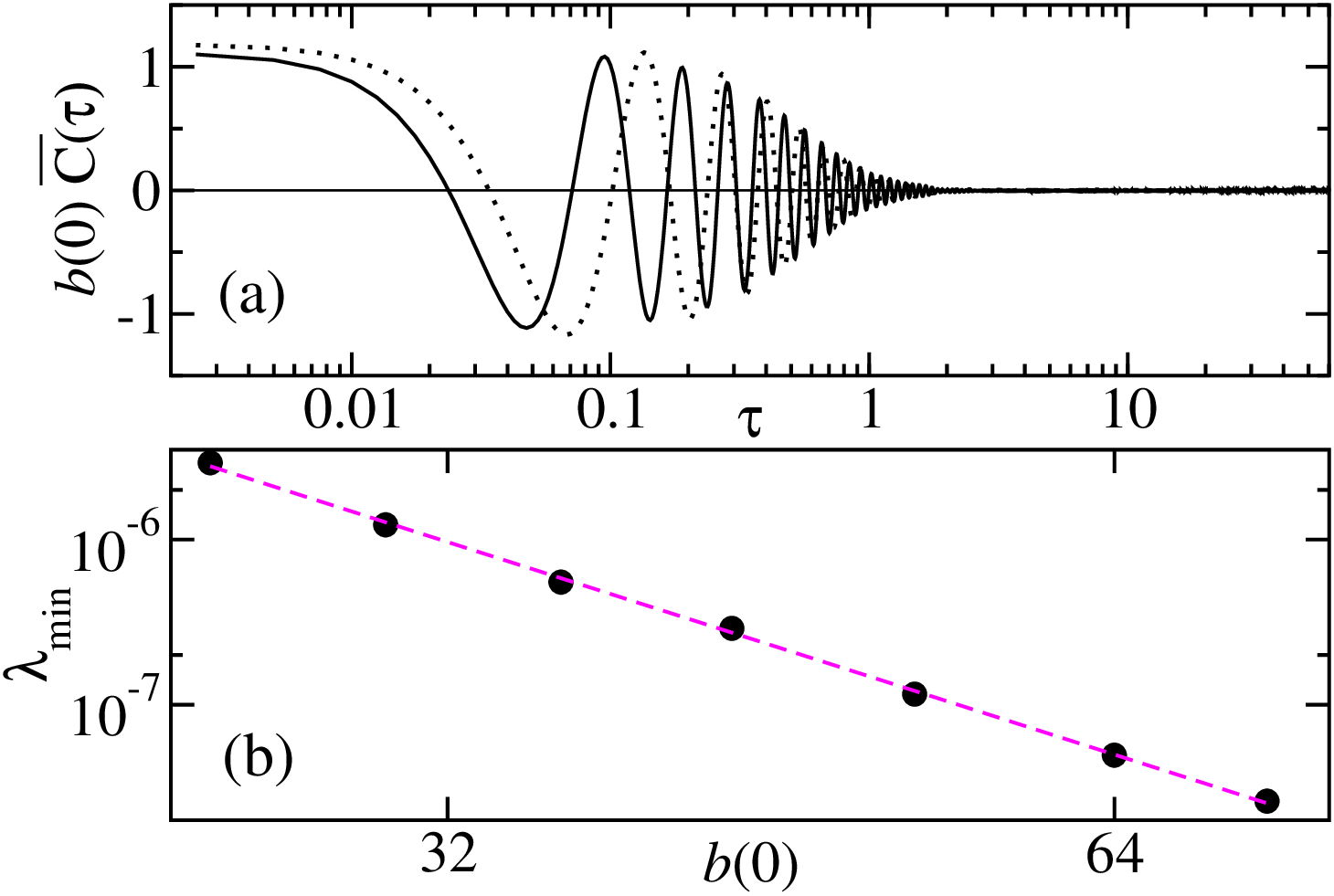}
\end{center}
\caption{(a) Average correlation for two different initial masses: $b(0)=25$ (solid curve) and $b(0)=35$ 
(dotted curve);
(b) PCA analysis: the smallest eigenvalue of $K_{ij}$ is plotted versus the mass $b(0)$.
}
\label{fig_corr}
\end{figure}

We start with the pseudo-stationary periods. Since they are relatively long, it makes sense to compute the correlation
$C(\tau) = \langle b(t+\tau)b(t)\rangle -\langle b(t)\rangle^2$, 
where the angular brackets denote a time average (computed over an interval of order $10^4$).
In Fig.~\ref{fig_corr}a, we report the correlation $\overline{C}(\tau)$, obtained by further averaging over
50 different initial conditions, all with the same $b(0)$. There we see that the stochastic-like dynamics of the
background quickly damps the correlations on a time scale $\tau \approx 1$. The comparison between the outcome
for mass 25 and 35 shows also that the amplitude of the correlations scales approximately as $1/b(0)$. 
Memory of the initial condition is, however, not entirely lost: the sample-to-sample fluctuations of $C(0)$
(upon changing the initial configuration of the background) are indeed 10 times larger than the statistical error
affecting $C(0)$ as obtained from a pure time average.

In order to shed further light, we have implemented the principal component analysis 
(PCA)~\cite{jolliffe2002principal}. 
Given a generic lattice configuration, we consider the three variables $[z_{-1},z_0,z_1]$ 
and rotate them until the breather variable $z_0$ is real and positive (the DNLS evolution is invariant 
under a homogeneous phase shift). The resulting state $[\tilde z_{-1},\tilde z_0,\tilde z_1]$ 
can be thereby parametrized by five real variables,
$[u_1,u_2,u_3,u_4,u_5]\equiv [\mathcal{R}(\tilde z_{-1}),\mathcal{I}(\tilde z_{-1}),\tilde z_0,
\mathcal{R}(\tilde z_{1}),\mathcal{I}(\tilde z_{1})]$.
Given an ensemble of such quintuplets, the correlation matrix 
$K_{ij} = \langle u_i u_j \rangle -\langle u_i\rangle\langle u_j\rangle$, is determined by averaging over time.
The real positive eigenvalues $\lambda_m$ of $K_{ij}$ correspond to the variance of the underlying quasi-stationary distribution along the so-called principal axes. It turns out that while four out of the five eigenvalues are close to 1, independently of the mass $b(0)$, the last eigenvalue $\lambda_{min}$ is very small and decreases upon increasing $b(0)$. From the data reported in Fig.~\ref{fig_corr}b, it follows that $\lambda_{min} \approx b(0)^{-4}$, meaning that the quasi-stationary regime unfolds within a thin flat manifold of thickness $\xi \approx b(0)^{-2}$. 

This result can be taken as evidence of a quasi-conserved quantity $\Qpca$ and explains the sample-to-sample
fluctuations of the correlation.
In fact, the variability of $C(\tau)$ signifies that the trajectory explores
different portions of the phase space (characterized by different values of $\Qpca$),
depending on the initial configuration of the background.
A linear, approximate expression of the pseudo-invariant manifold can be obtained 
from the eigendirection corresponding to $\lambda_{min}$.
In terms of the canonical variables, defined by 
$z_0 = \sqrt{b}\exp(i\phi_0)$ and $z_k=(p_k +iq_k)/\sqrt{2}$ for $k\ne 0$,
\begin{equation}
\Qpca = \sqrt{b} + c [(p_1+p_{-1}) \cos \phi_0 + (q_1+q_{-1}) \sin \phi_0] \, ,
\label{eq:q1}
\end{equation}
where $c$ is a small quantity, which decreases upon increasing $b(0)$.

The quasi-conservation law suggests the existence of an adiabatic invariant (AI)~\cite{bookvariational}.
This is not surprising, since AIs typically arise in the presence of two widely separated time scales,
here associated to the background dynamics and the rapid rotation of the breather.
On a more quantitative level, we can implement a perturbative approach, starting from the definition of
the smallness parameter $\varepsilon^2 =1/b(0)$. 

\begin{figure}
\begin{center}
\includegraphics[width=0.45\textwidth,clip ]{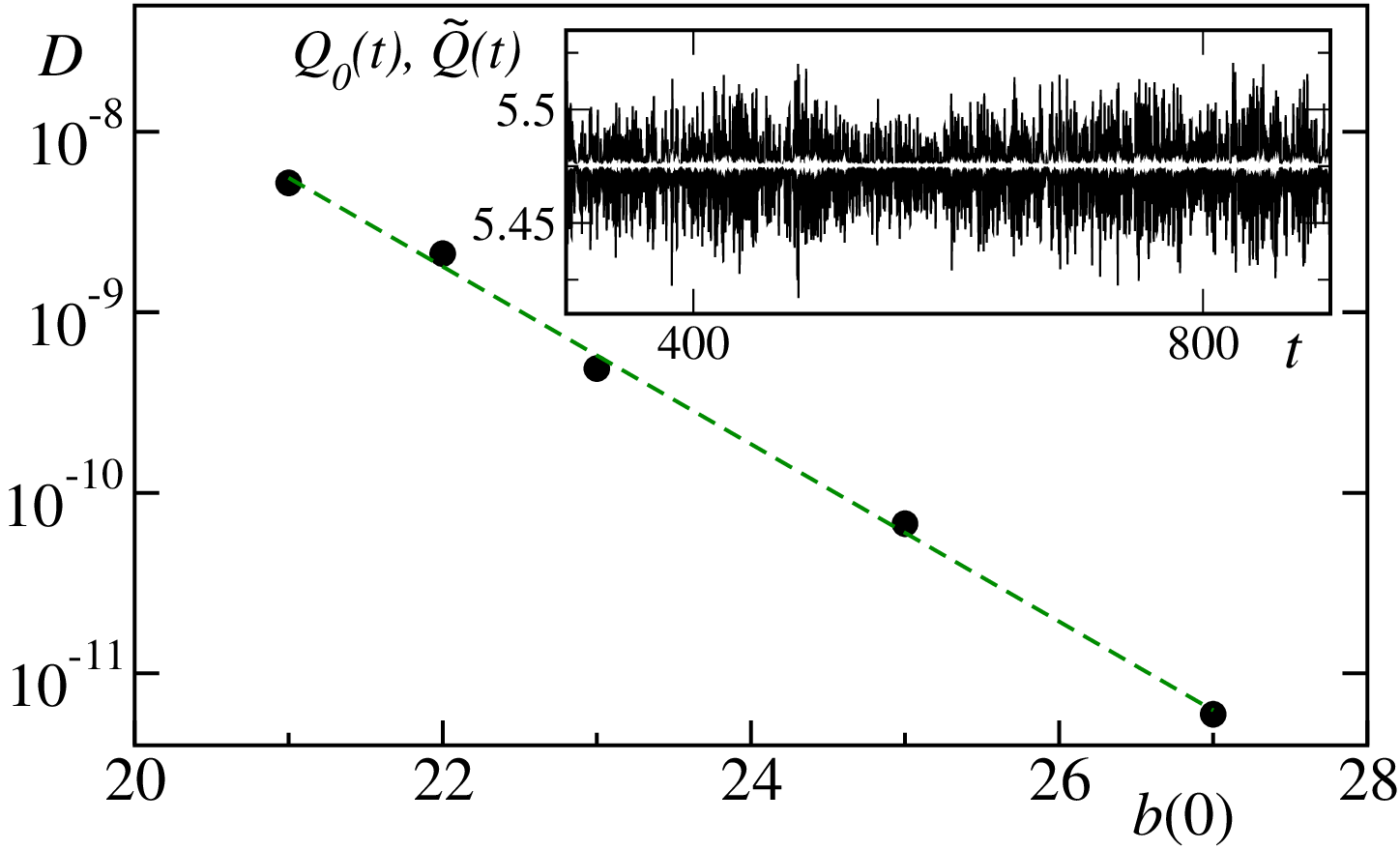}
\end{center}
\caption{Diffusion coefficient $D$ as a function of the initial breather mass.
The inset shows the evolution of $\tilde Q(t)$ (white line) compared to that of the breather amplitude $Q_0(t)$ (black line).}
\label{fig_diff_full}
\end{figure}

If we rescale the breather mass, introducing $B(t)\equiv b(t)/b(0)$, all variables are of order one and the
Hamiltonian takes the form 
$H=\varepsilon^{-4}[H_0 + \varepsilon^3 H_3 + \varepsilon^4 H_4]$, with
\bea
H_0 &=& B^2 \\
H_3 &=& \sqrt{2B}[(p_1+p_{-1}) \cos \phi_0 + (q_1+q_{-1}) \sin \phi_0] \\
H_4 &=& \frac{1}{4}(p_1^2+q_1^2)^2 +\frac{1}{4}(p_{-1}^2+q_{-1}^2)^2 + \ldots
\eea
An adiabatic invariant $Q=Q_0+\varepsilon Q_1 + \ldots$ can be thereby determined by 
imposing that the Poisson brackets $\{H,Q\}=0$ vanish.
At zero order, $\{ H_0,Q_0\}=0$ implies $Q_0=Q_0(B)$, which simply means that  arbitrarily tall breathers 
are decoupled from the background. The first correction arises at third order: 
$\{H_0,Q_3\} + \{H_3,Q_0\} = 0$ implies $Q_3 = (H_3/2B)(dQ_0/dB)$. 
In principle any choice of $Q_0(B)$ is possible but the PCA analysis suggests to select
$Q_0 = \sqrt{b(0) B}$.
In terms of unscaled variables, the truncated quantity $\tilde Q \equiv Q_0+\varepsilon^3Q_3$
is equal to $\Qpca$ once we set $c=1/(2\sqrt{2}b)$, see Eq.~(\ref{eq:q1}). The comparison between $Q_0$ and
$\tilde Q$ presented in the inset of Fig.~\ref{fig_diff_full} confirms that the latter quantity exhibits much
smaller fluctuations.

AIs are not exact conservation laws. Over sufficiently long time scales, large deviations are indeed
typically observed.
In the Klein-Gordon lattice, for instance, it has been proven that an AI may destabilize over exponentially
long time scales in the thermodynamic limit~\cite{Carati2012}.
This is true also in the DNLS equation. A first quantitative evidence is given by the diffusive behavior
of $\tilde Q$.
In practice, we have computed
$\Gamma(\tau) = \langle [\tilde Q^2(t+\tau)-\tilde Q^2(t)]^2 \rangle$ for a time 
$\tau$ long enough to see $\Gamma(\tau)$ growing linearly%
~\footnote{The reason of the inner square in the definition of $\Gamma$ is that we 
are interested in a mass-like variable such as $Q^2$.}.
The results for $D = \Gamma(\tau)/\tau$ are plotted in Fig.~\ref{fig_diff_full}~\cite{sup_mat02}, 
where the diffusion coefficient is shown to decrease exponentially with the breather mass, 
$D \approx \mathrm{e}^{-\gamma b(0)}$, with $\gamma = 1.13 \pm 0.09$.

During the laminar periods, the background is basically at equilibrium with a temperature 
and a chemical potential set by the external reservoir. 
Accordingly, the background itself can be interpreted as an effective thermal bath, which
interacts directly with the breather. 
Fluctuation-dissipation considerations then suggest that the interaction should be characterized by a diffusion
coefficient that can be identified with $D$, and a drift $v$, responsible for the eventual
absorption of the breather. In fact, for $T>0$ the equilibrium state of the DNLS is 
statistically homogeneous, with no breathers.
According to the same fluctuation-dissipation considerations, $v$ is expected to be proportional to 
$D/T$, as also confirmed by explicit calculations for a simple model~\cite{iubini17}.
As a result, in the absence of jumps, we expect that the lifetime of the breather should be
at least of the order of $b(0)/v \approx b(0) \mathrm{e}^{\gamma b(0)}$. 
Notice that $\gamma$ is slightly larger than the direct estimate $\alpha$.
We can conclude that the laminar-phase dynamics is compatible with the exponential growth of the 
breather lifetime~\footnote{We have not been able to estimate directly
$v$, presumably since it is smaller than $D$, because of the $1/T$ multiplicative factor.}.

\begin{figure}
\begin{center}
\includegraphics[width=0.45\textwidth,clip ]{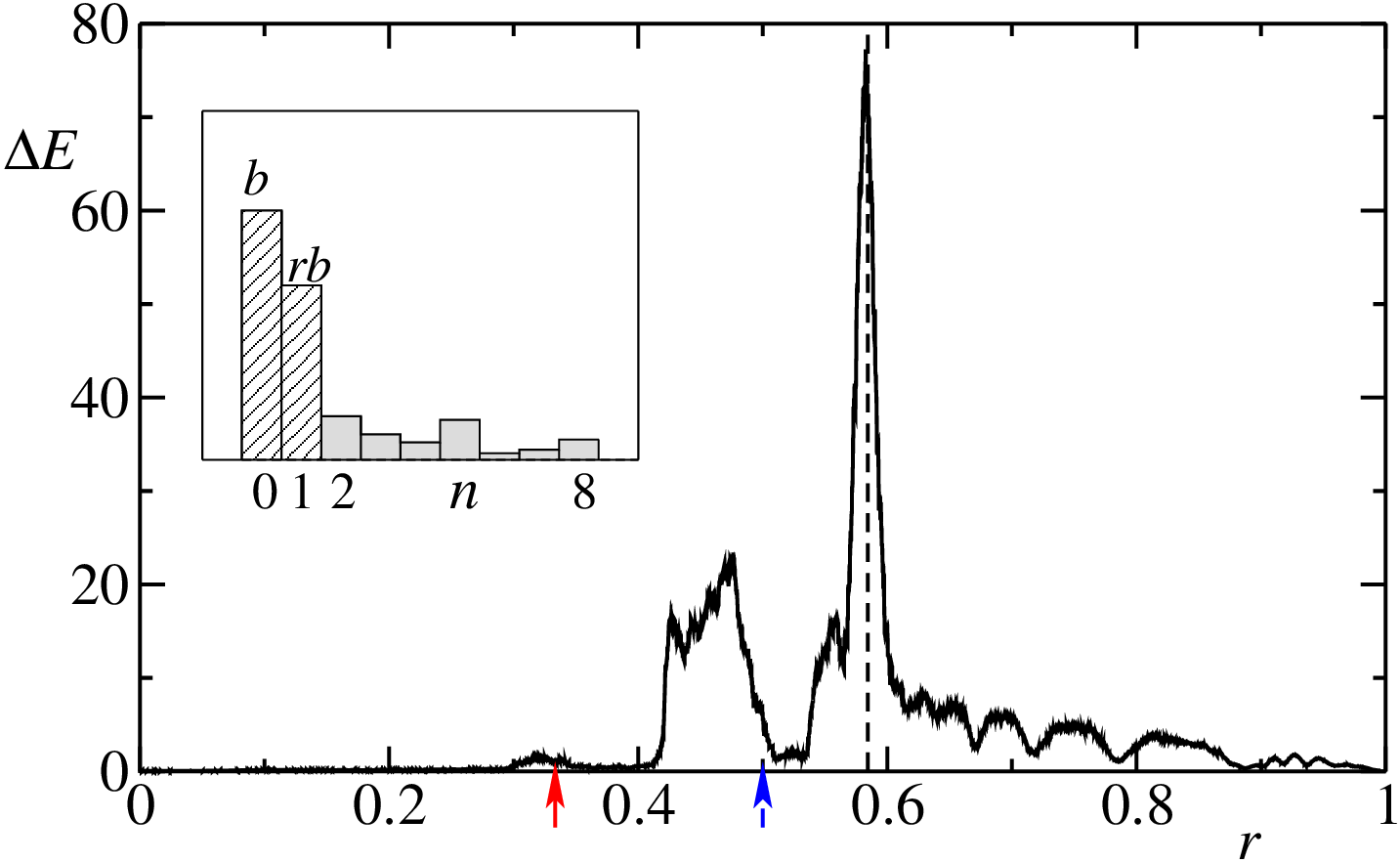}
\end{center}
\caption{Average jumps of breather energy, $\Delta E$, as a function of the initial mass ratio $r$.  
The inset depicts the geometry of the system.
The highest peak corresponds to the analytical value $r_d$ (dashed line).
The resonance values $r_3$ and $r_2$ are indicated by the arrows.
}
\label{transition_peaks}
\end{figure}

Where does the slowness of the relaxation process come from?
One cannot, strictly speaking, invoke a gap in the spectrum between the frequency of the breather
and the surrounding waves, because the nonlinear nature of the background implies a
broadband spectrum. One might still naively trace back the slowness to the small amplitude of 
the background power spectrum at the breather frequency.
However, this is not the case; we have tested that a pure and simple interaction of the breather with 
a stochastic process characterized by the same  power spectrum of the
background gives a far much faster relaxation, so that it is necessary to account for the effects
of the breather on the background itself.

Mathematically speaking, one reason for the non perfect invariance of an AI is the lack of convergence
of the underlying expansion, when the Hamiltonian perturbation terms (here $\varepsilon^3H_3$ and 
$\varepsilon^4 H_4$) are non negligible.
This may indeed happen whenever the amplitude in the breather neighboring sites is occasionally very large.
In fact, preliminary simulations show that such events induce $\tilde Q$-jumps and thereby terminate the long
laminar periods.

In order to quantify this phenomenon, we have computed $\Delta E=\overline{|\Delta b^2|}$, 
where $b^2(t)$ is the breather energy, while
$\Delta E$ indicates its variation after 50 time units~\cite{sup_mat02}.
The breather, of initial mass $b(0)=36$, is set on the left boundary of the chain (no coupling with
the left neighbor), while the rightmost 8th site is thermalized at $T=10,\mu=-6.4$.
Finally, the mass of site $1$ is set equal to $b_1(0) = r b(0)$.
A plot of $\Delta E$ as a function of $r$ shows a clear peak (see Fig.~\ref{transition_peaks})
at $r_d = (\sqrt{b(0)} - \sqrt{2})^2/b(0)$ (see the vertical line); it corresponds to the 
activation threshold for the formation of a symmetric, two-site, localized structure.
Analogously to the breather, the localized-structure is weakly coupled with the background 
when the amplitude of both sites is large enough. Therefore, it can be approximately treated as
a dimer configuration with open boundary conditions and, from now on, we refer to it simply
as to a {\it dimer}.

Symmetric dimers are characterized by periodic oscillations of the mass between 
$b_{min}$ and $b_{max}$~\cite{Kenkre1986}.
Additionally, for a given $b_{tot}= b_{min}+b_{max}$, there exists a minimal $b_{min}$ for
the oscillations to self-sustain. This is the origin of the above mentioned threshold $r_d$
for the ratio $r$.
A symmetric dimer is relatively stable (though much less than the single breather),
but eventually collapses onto a single breather, which can possibly hop onto a neighboring site,
a phenomenon that is indeed observed in numerical simulations (see, e.g., the inset of Fig.~\ref{fig_brt_relax}).
As shown in Fig.~\ref{fig_lifetime}, the average first hopping time $\tau_h$ (open diamonds) 
increases exponentially with $b(0)$ (with a rate $\alpha_h=1.00\pm 0.04$), revealing that dimer formation and jumps 
seem to be related to the observed exponentially-long breather lifetimes.

A semi-quantitative estimate of $\tau_h$ can be obtained by approximating it with the (average) time $\tau_\theta$ 
required for a background fluctuation to reach the threshold $\theta= r_d b$; $\tau_\theta$ roughly corresponds
to the inverse of the probability $P(\theta)$ to observe the mass $\theta$ at equilibrium.
From \cite{Rasmussen2000}, we know that for large temperature $T=1/\beta$, 
$P(\theta) = \sqrt{4\beta/\pi}\mathrm{e}^{-\beta(\theta^2-\mu \theta+\mu^2/4)}/[1+\mbox{erf}(\mu\sqrt{\beta}/2)]$.
In the limit of large $b$, $\tau_\theta\approx \exp(r_d^2\beta b^2)$, i.e.
this rough argument suggests that first-hopping time might even grow super-exponentially with $b$.
Unfortunately, this prediction is not fully quantitative, as the probability density in the
breather nearest neighbors is affected by the breather itself and is only approximately equal to
the equilibrium distribution.

The onset of symmetric dimers is not the only means to transfer mass out of a breather. By looking at 
Fig.~\ref{transition_peaks}, one can see additional peaks, which approximately 
coincide with resonances, where the frequency on site $1$ is equal to $1/3$ ($r_3$) or $1/2$ ($r_2$) of 
the breather frequency. 
Although the single resonance events are not so effective, they are much more frequent than dimer-formation events
and might be relevant for ``killing'' the adiabatic invariant~\cite{neishtadt2006}.
The relative weight of the two relaxation mechanisms (dimers and resonances) with varying $b(0)$ is still unclear.

All of our studies consistently give evidence of an exponentially slow relaxational dynamics.
The origin of such freezing process is very different from the arrest mechanisms typically
encountered in statistical mechanics: it has a purely dynamical origin, being enforced by the
existence of an adiabatic invariant (AI).
For $T>0$, the AI neutralizes entropic forces, preventing {\it de facto}  
a macroscopic relaxation as soon as one tall breather is contained in the initial configuration.
Indeed, in the presence of an exponentially weak effective breather-background interaction, 
breather condensation proceeds through a practically unobservable logarithmic coarsening~\cite{iubini17}.
In the negative-temperature region, the same mechanism prevents breather growth, thereby
``stabilizing'' a fairly homogeneous chaotic non-ergodic dynamics, as suggested by
recent direct numerical simulations~\cite{iubini2013,mithun18}. 

A final and more detailed understanding of the problem requires on the one hand including higher-orders in the perturbation analysis
to estimate the convergence properties of the AI, on the other hand identifying and describing the most effective perturbations
responsible for the sporadic mass transfer.
A deep understanding of slow relaxation phenomena will actually be important to analyze frozen dynamics in 
driven systems, both in DNLS itself~\cite{wall17} and in systems of rotors
where the conductivity is exponentially small upon increasing the temperature~\cite{pino16}.
This phenomenon has been shown to be related to the emergence of many-body
quantum localization and ergodicity breaking~\cite{pino16,Cuneo_3rotators2015,cuneo2016_4r,cuneo2017relaxation}.

\begin{acknowledgments}
We thank S. Lepri and R. Livi for a critical reading of the manuscript.
S.I and A.P. thank S. Paleari for useful discussions of adiabatic invariants.
S.I. acknowledges support from Progetto di Ricerca Dipartimentale BIRD173122/17.
L.C. and G.-L.O thank the Carnegie Trust for the Universities of Scotland for financial support.
\end{acknowledgments}

%

\end{document}


\renewcommand{\thepage}{S\arabic{page}}  
\renewcommand{\thesection}{S\arabic{section}}   
\renewcommand{\thetable}{S\arabic{table}}   
\renewcommand{\thefigure}{S\arabic{figure}}
\renewcommand{\theequation}{S\arabic{equation}}
\renewcommand{\bibnumfmt}[1]{[S#1]}
\renewcommand{\citenumfont}[1]{S#1}

\title{Supplemental material for\\``Dynamical freezing of relaxation to equilibrium"}

\author{Stefano Iubini}

\author{Liviu Chirondojan}

\author{Gian-Luca Oppo}

\author{Antonio Politi}

\author{Paolo Politi}

\maketitle

\centerline{\large\bf Details of computational analysis}
\vskip 1cm

\paragraph{Numerical evolution of the DNLS dynamics in the presence of energy and mass reservoirs -} Given the need to run long simulations, we defined an optimal set-up to minimize
the computation time. As a reference setup,  we have chosen to simulate DNLS chains of length $2N_0+1$
with a breather sitting in the middle and both chain ends attached to suitable heat baths (see below).

It is desirable to choose $N_0$ as short as possible, but not too short otherwise 
the overall scenario is strongly affected by boundary layers which emerge in the
vicinity of the heat baths. We have verified that for $N_0\ge 9$ such effects are negligible
and the lifetime is independent of $N_0$. Two different integration schemes have been implemented:

(i) Langevin-type thermal baths~\cite{ILLP2013}  together with a standard fourth-order Runge-Kutta algorithm~\cite{press2007numerical} (and a sufficiently
small integration time step even down to $10^{-5}$ time units so as to follow the fast rotation of the breather);

(ii) Monte Carlo thermal baths~\cite{iubini2012} with a symplectic fourth-order Yoshida algorithm~\cite{yoshida1990} (minimum time-step $10^{-3}$ time units).

For scheme (i), the explicit Langevin equation (specified for the last lattice site) reads
\begin{equation}
\hspace{-1.cm} i \dot z_{N_0}= (1+i \Gamma)\left[-2|z_{N_0}|^2 z_{N_0}  -z_{N_0-1} \right]  
 +i\Gamma \mu z_{N_0}+ \sqrt{\Gamma T} \, \eta(t) \quad  \; ,
 \end{equation}
where $\eta(t)$ is a complex Gaussian white noise with zero mean and unit variance and $\Gamma$ is the bath coupling parameter. 
Without any loss of generality,  we have chosen $\Gamma=1$. 

For scheme (ii), according to~\cite{iubini2012}, the two Monte Carlo reservoirs interact with the DNLS chain at random times whose separations are independent and distributed uniformly within the interval $[t_{\rm min},t_{\rm max}]$. We have chosen $t_{\rm min}=0.4$ and
$t_{\rm max}=2$. 
We have verified that the two numerical approaches are consistent with each other. 

\vspace{0.5 cm}
\paragraph{Figure 4 -}The results for the diffusion coefficient have been obtained by scanning the entire breather
evolution and discarding all the time intervals where the mass of  either neighbor site is
larger than 8, to exclude the occasional jumps caused by sudden increase of the amplitude in
the neighboring sites.

\vspace{0.5 cm}
\paragraph{Figure 5 -}The sightly different setup with the breather placed on one boundary of the DNLS chain is chosen
in order to cleanly study the effect of just one neighboring background site with an anomalously large mass fluctuation.
The results shown in Fig. 5 refer to an ensemble of initial conditions where the phase difference $\phi_2-\phi_1$ is set to $\pi$.
This condition ensures the largest coupling with the breather. 